\documentclass[conference]{IEEEtran}
\usepackage[pdftex]{graphicx}
\usepackage{amssymb}
\usepackage{bm}
\usepackage{amsmath}
\usepackage{pgfplots}
\usepackage{balance}
\usepackage{epstopdf}
\usepackage{algpseudocode}
\usepackage{algorithm,algpseudocode}
\usepackage{algorithmicx}
\usepackage{dblfloatfix} 
\usepackage[noadjust]{cite}
\usepackage{booktabs} 
\pgfplotsset{compat=newest}
\pgfplotsset{plot coordinates/math parser=false}
\newlength\figureheight
\newlength\figurewidth

\usepackage{color,soul}
\IEEEoverridecommandlockouts

\ifCLASSOPTIONcompsoc
    \usepackage[caption=false, font=normalsize, labelfont=sf, textfont=sf, subrefformat=parens, labelformat=parens]{subfig}
\else
\usepackage[caption=false, font=footnotesize, subrefformat=parens, labelformat=parens]{subfig}
\fi
  
\usepackage{amsmath,amssymb}
\usepackage{mathtools} 
\usepackage{xspace}

\newcommand{\Figref}[1]{Fig.~\ref{#1}}

\newcommand{\boundellipse}[3]
{(#1) ellipse [x radius=#2,y radius=#3]
}

\newcommand{\EX}[1]{\mathsf{E}\left\{{#1}\right\}}

\newcommand{\Pd}{\rho_{\mathrm{d}}}

\makeatletter
\def\@setsize#1#2#3#4{
    \@nomath#1
    \let\@currsize#1
    \baselineskip #2
    \baselineskip \baselinestretch\baselineskip
    \parskip \baselinestretch\parskip
    \setbox\strutbox \hbox{
        \vrule height.7\baselineskip
            depth.3\baselineskip
            width\z@}
    \skip\footins \baselinestretch\skip\footins
    \normalbaselineskip\baselineskip#3#4}
\makeatother

\makeatletter
\newcommand{\setstretch}[1]{
    \def\baselinestretch{#1}%
    \@currsize
    }
\makeatother



\begin{document}

\begin{figure*}[t!]
\normalsize
Paper published in ICC 2019 -- 2019 IEEE International Conference on Communications (ICC). \\Added to IEEE Xplore: July 15, 2019. DOI: 10.1109/ICC.2019.8761828. 

\

\textcopyright~2019 IEEE. Personal use of this material is permitted.  Permission from IEEE must be obtained for all other uses, in any current or future media, including reprinting/republishing this material for advertising or promotional purposes, creating new collective works, for resale or redistribution to servers or lists, or reuse of any copyrighted component of this work in other works.
\vspace{17cm}
\end{figure*}

\newpage

\title{Scalability Aspects of Cell-Free Massive MIMO}
\author{\IEEEauthorblockN{Giovanni Interdonato$^{*\dagger}$, P{\aa}l Frenger$^{*}$ and Erik G. Larsson$^\dagger$}
\IEEEauthorblockA{$^*$Ericsson Research, 583 30 Link\"oping, Sweden\\
$^\dagger$Department of Electrical Engineering (ISY), Link\"oping University, 581 83 Link\"oping, Sweden\\
\{giovanni.interdonato, erik.g.larsson\}@liu.se, pal.frenger@ericsson.com\thanks{This paper was supported by the European Union's Horizon 2020 research
and innovation programme under grant agreement No 641985 (5Gwireless), and ELLIIT.}}}
\maketitle

\begin{abstract}
Ubiquitous cell-free massive MIMO (multiple-input multiple-output) combines massive MIMO technology and user-centric transmission in a  distributed architecture. 
All the access points (APs) in the network cooperate to jointly and coherently serve a smaller number of users in the same time-frequency resource. 
However, this coordination needs significant amounts of control signalling which introduces additional overhead, while data co-processing increases the back/front-haul requirements. 
Hence, the notion that the ``whole world'' could constitute one network, and that all APs would act as a single base station, is not scalable. 
In this study, we address some system scalability aspects of cell-free massive MIMO that have been neglected in literature until now. In particular, we propose and evaluate a solution related to data processing, network topology and power control. Results indicate that our proposed framework achieves full scalability at the cost of a modest performance loss compared to the  canonical  form of cell-free massive MIMO.  
\end{abstract}

\begin{IEEEkeywords}
Cell-free Massive MIMO, distributed wireless system, power control, spectral efficiency, system scalability. 
\end{IEEEkeywords}

\section{Introduction}
Coordinated distributed wireless systems~\cite{Zhou2003a} leverage signal co-processing at multiple access points (APs) to guarantee high connectivity, reduce inter-cell interference and improve the user experience. 
Connectivity is enhanced thanks to the shorter AP-to-user distance; the joint coherent transmission from geographically distributed APs yields macro-diversity gain; and the coordination enables APs
to select transmit strategies jointly (by sharing channel state information)  in order to reduce inter-cell interference.

Ubiquitous cell-free massive MIMO~\cite{Nayebi2015a,Ngo2017b,Interdonato2018a}, where MIMO stands for multiple-input multiple-output, relies on these principles.
However, it differs from prior coordinated distributed wireless systems such as network MIMO~\cite{Venkatesan2007a,Caire2010b}, virtual MIMO~\cite{Feng2013c}, multi-cell MIMO cooperative networks~\cite{Gesbert2010a} and coordinated multipoint with joint transmission (CoMP-JT)~\cite{Boldi2011a,Marsch2011a,irmer2011coordinated}.
The performance of all the aforementioned frameworks, in their canonical forms, are limited by two factors: $(i)$ the inter-cell (out-of-cluster) interference inherent in the static cell-centric transmission design~\cite{Lozano2013a}; $(ii)$ the large amount of control signalling, channel state information (CSI) and data exchange, required for the coordination, that increases the back/front-hauling requirements and the overhead. These become more significant as the number of APs grows.
  
Cell-free massive MIMO combines the benefits derived from using time division duplex (TDD) massive MIMO technology~\cite{Marzetta2016a} and user-centric transmission. 
The TDD massive MIMO operation brings high spectral and energy efficiency, and enables  substantial reductions of the estimation overhead as the uplink estimates can be reused in the downlink for precoding by exploiting (after calibration)    channel reciprocity. 
The conventional operation consisting in estimating the downlink channel followed by the user feedback transmission can be conveniently avoided. Hence, the estimation overhead scales with the number of users rather than the number of APs. 
Moreover, the massive number of serving APs brings a two-fold benefit: it introduces additional macro-diversity and it reduces the multi-user interference, thanks to the \textit{favorable propagation}~\cite{Marzetta2016a} phenomenon. 
The inter-cell interference is effectively suppressed by the user-centric data transmission design: each user equipment (UE) is surrounded by serving APs, hence, it experiences no cell boundaries. 

Although many aspects of a cell-free massive MIMO network have been deeply studied in the literature, system scalability has not been fully addressed until now. In the canonical case~\cite{Nayebi2015a}, a cell-free massive MIMO network comprises many distributed APs simultaneously serving a smaller number of UEs. 
All the APs are connected to one central processing unit (CPU) which is responsible for coordination and data processing. 
Hence, from the UE perspective, the entire network acts as an infinitely large single cell, served by a single base station comprising all the APs and the CPU. 
Such a framework is certainly unrealistic in practice, and the notion that data from all the APs would be processed coherently is not scalable.

\textbf{Contributions:} In this study, we propose a scalable cell-free massive MIMO framework, specifically considering data transmission strategies  and power control. 
Unlike prior studies, we consider a realistic scenario where multiple CPUs serve disjoint clusters of APs. 
Finally, the spectral efficiency provided by the proposed scheme is compared with the canonical cell-free massive MIMO, and conventional (cell-centric) CoMP-JT.

\section{The Scalability Problem}

Compared to prior coordinated distributed wireless systems, canonical cell-free massive MIMO is a step forward in terms of system scalability.
As mentioned earlier, the TDD operation allows to make the estimation overhead independent of the number of APs. 
In addition, the use of conjugate beamforming (i.e., maximum-ratio transmission, MRT), which has been especially advocated in the literature, enables a simple and fully distributed processing. 

Let $M$, $K$ the number of APs and UEs, respectively. The data signal sent by AP $m$ to all UEs, using MRT, is given by
\begin{equation} \label{eq:transmission-model}
x_m = \sqrt{\Pd}\sum\nolimits_{k=1}^K \sqrt{\eta_{mk}} \hat{g}^\ast_{mk} q_k,
\end{equation} 
where $q_k$ is the unit-power data symbol intended for the $k$th UE, $\Pd$ is the normalized transmit signal-to-noise ratio related to the data symbol (i.e., the radiated power over the power of the noise figure). The term $\hat{g}^\ast_{mk}$ represents the precoding factor, namely the conjugate of the channel estimate between AP $m$ and UE $k$. The power spent by AP $m$ on the service of UE $k$ is parametrized in terms of a power control coefficient $\eta_{mk}$, where $0 \leq \eta_{mk} \leq 1$~\cite{Ngo2017b}.   

With MRT, the precoders are determined at each APs by using only local CSI, thus no CSI and information about precoders are exchanged over the front-haul network between APs and CPU. Hence, precoding is also scalable. 

However, there are at least three main scalability issues:
\begin{itemize}
\item \textit{Data processing.} Data destined for every UE in the network would have to be sent from the CPU to every AP. This would render the computational complexity at each AP unsustainable;
\item \textit{Network topology.} The complexity of the interconnect at the CPU does not scale as the CPU will need one connection to each AP in the network;
\item \textit{Power control}. The calculation of the power control coefficients does not scale, even taking computational issues aside. This aspect is discussed in detail next.
\end{itemize}   

\subsection{Is Power Control Really Scalable?}
Each AP has a transmission power constraint related to $\Pd$, i.e., the per-AP power constraint is given by 
\begin{equation}
\EX{|x_m|^2} \leq \Pd, \quad \forall m,
\end{equation}
which can be expressed as,
\begin{equation}
\label{eq:power-constraint}
\sum\nolimits_{k=1}^K \eta_{mk} \gamma_{mk} \leq 1, \quad \forall m,
\end{equation}
where $\gamma_{mk}$ is the mean-square of the channel estimate, i.e., $\EX{|\hat{g}_{mk}|^2} = \gamma_{mk}$. It is proportional to the mean-square of the effective channel $\beta_{mk}$~\cite{Ngo2017b}.

For given power control coefficients, analytical capacity lower bounds (``achievable rates'') exist that quantify performance given some pre-determined path-loss and fading model. 
The power control coefficients are functions only of the long-term channel statistics and must be computed centrally. Hence, they need to be sent by the CPU to all the APs. 

Algorithms for the optimal selection of $\{\eta_{mk}\}$ are available. Specifically, max-min fairness power control, that ensures that every UE in the network obtains the same quality of service (rate), is possible (though computationally very demanding) through the use of convex optimization tools~\cite{Ngo2017b}. 
However, the power control coefficient associated with some UE $k$ and some AP $m$ depends on the channel statistics of UE-AP pairs very far away. This ``butterfly effect'' entangles the power control coefficients across the whole network.

Simpler effective policies, proposed in~\cite{Ngo2017b,Nayebi2017a}, enable distributed computation of the power control coefficients at the cost of reduced performance. 
In the distributed power control case, there is no power control coefficient exchange over the front-haul network, and only the power control coefficients associated to the same AP are entangled each other (local ``butterfly effect'') to satisfy the per-AP power constraint in~\eqref{eq:power-constraint}.    

\subsection{User-centric vs Cell-centric Clustering}
The basic way to attempt to address the scalability problem consists in deploying clusters of APs and to confine the signal co-processing within the cluster. 
In literature, there are two suggested approaches: $(i)$ \textit{cell-centric clustering}, which consists in deploying fixed disjoint clusters of APs where the APs in a cluster serve only the UEs residing in their joint coverage area; $(ii)$ \textit{user-centric clustering}, which consists in deploying dynamic (possibly partially overlapped) clusters of APs based on the needs of each served UE. 

Cell-centric clustering constitutes the canonical method to group cooperating APs. 
Each cluster is served by one CPU, and the APs connected to a given CPU will form a cluster. 
These clusters will either mutually interfere, or they will have to cooperate through coherent transmission, which brings back the scalability problem. 
In such a system there is no coherent cooperation on data, or cooperation on power control, between the CPUs.
Although this system is fully scalable, it suffers 
from poor performance (it will be addressed later). 
Network MIMO, CoMP-JT and multi-cell MIMO cooperative network are conventionally implemented in a cell-centric fashion.    

Conversely, the idea behind the user-centric clustering is that, for each UE, only a small number of APs should participate in the service of that UE. Effectively each UE is served by a cluster of near-by APs.
One can view user-centric transmission as a special case of the common cell-free massive MIMO setup where all power control coefficients $\{\eta_{mk}\}$ of a given UE $k$ are constrained to be zero, except for those associated with the closest APs.
The user-centric approach enables to suppress the inter-cell interference thus it performs better than the cell-centric approach. 
However, this concept fundamentally does not solve the network topology problem described earlier, i.e., that all APs must be connected to a CPU. 
In addition, it requires more control signalling to dynamically form user-specific clusters ``on demand''. 
User-centric approach has been subsequently introduced in multi-cell cooperative network~\cite{Papadogiannis2008a}, CoMP-JT~\cite{Baracca2012b}, cooperative small cells, under the name of \textit{cover-shifts}~\cite{Jungnickel2014b}, and C-RAN\footnote{Cloud Radio Access Network is an architecture that moves the baseband processing from the APs to ``the cloud''. 
Today, C-RAN is mainly used to implement conventional cellular systems, but in the future it might be used to implement a cell-free massive MIMO system.}~\cite{Pan2018b,Yuan2017c}.         
\setcounter{figure}{1}     
\begin{figure*}[!b]
\hrule
\vspace*{4pt}
\centering
\includegraphics[width=\linewidth]{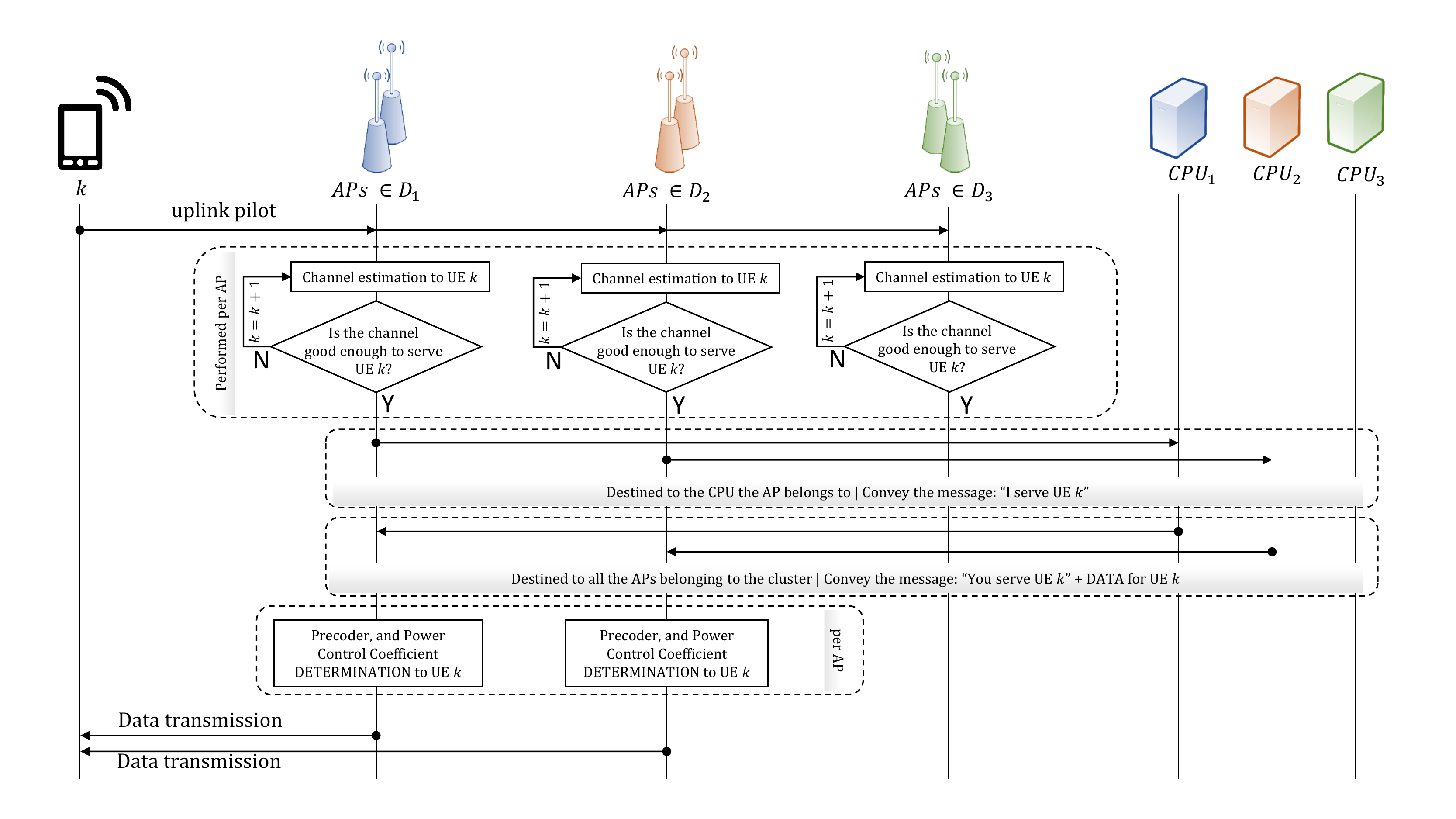}
\caption{Signalling diagram depicting the operation of the proposed cell-free massive MIMO framework.}
\label{fig:signaling-diagram}
\end{figure*}
       
\section{Proposed Solution}
In this section, we describe our proposed fully scalable and distributed cell-free massive MIMO framework.
Distributed resource allocation problems for cooperative beamforming systems have been considered in~\cite[Section 4.2]{Bjornson2013d}. However, to our knowledge no scalable solutions for power control and precoding are available that do not rely on perfect CSI assumptions. 

The APs are grouped into $N$ pre-determined cell-centric clusters $D_1, \ldots, D_N$.
Each cluster in turn is connected to one CPU. The CPUs are interconnected but operate autonomously. 
It is assumed that a global phase reference is shared.
Each UE is receiving service from one or a few cell-centric clusters. 
Let $B_k$ be the number of cell-centric clusters that participate in the service of the $k$th UE (typically this will be a small number). 
We denote these serving cell-centric clusters by $D_{k1},\ldots, D_{kB_k}$.
To select said clusters, the user-centric concept is applied. 
Specifically, for the $k$th UE, the $L_k$ selected APs, i.e., $\text{AP}_{k1},\ldots, \text{AP}_{kL_k}$, that form the user-centric cluster are identified, as shown in the example in \Figref{fig:proposed-system}.
\setcounter{figure}{0} 
\begin{figure}[!t]
\centering
\includegraphics[width=\columnwidth]{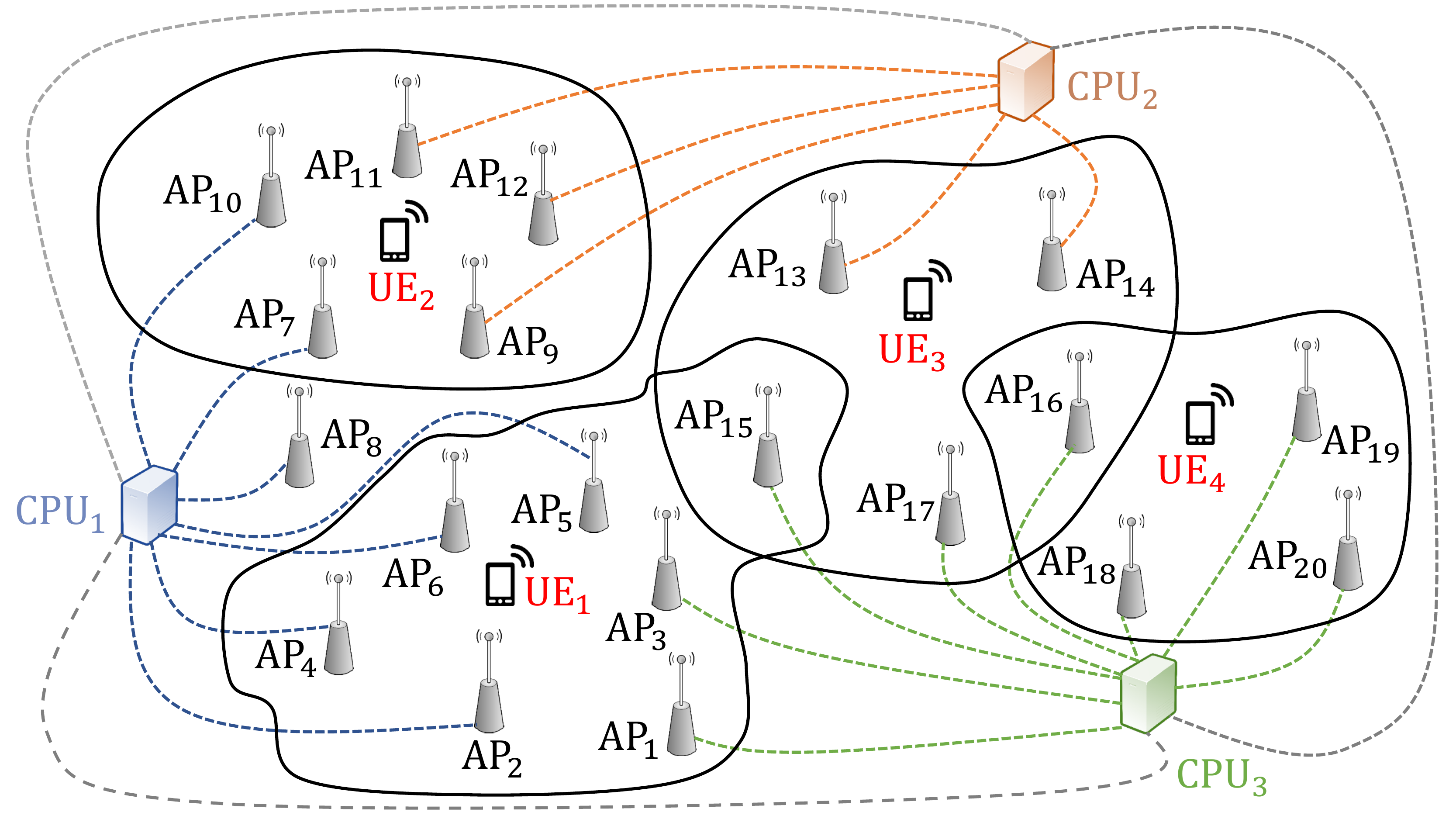}
\caption{Example of cell-free massive MIMO system with user-centric clustering and multiple interconnected CPUs. Different colors for CPUs and front-haul networks correspond to different cell-centric clusters.}
\label{fig:proposed-system}
\end{figure}
\setcounter{figure}{2} 
These APs are selected according to some metric, for example distance, or channel quality. 
The clusters involved by the selected APs define the serving cell-centric clusters. Hence, the data to UE $k$ is distributed only to the involved CPUs:  $\text{CPU}_{k1}, \ldots, \text{CPU}_{kB_k}$.

As shown in \Figref{fig:proposed-system}, a user-centric cluster might include APs belonging to different cell-centric clusters. 
Hence, UE $k$ is served by all the APs of the selected cell-centric clusters. 
For instance, in \Figref{fig:proposed-system}, $\text{UE}_2$ is served by all the APs managed by CPU$_1$ and CPU$_2$, that is cell-centric cluster $D_1$ and $D_2$.

Power control is applied independently in each APs. To satisfy~\eqref{eq:power-constraint} with equality, the power control coefficients are set as follows
\begin{equation}
\eta_{mk}\!=\! 
\begin{cases}
\frac{f(\mathcal{G}_{mk})}{\sum\limits_{k^\prime \in T_m}\! \gamma_{mk^\prime} f(\mathcal{G}_{mk^\prime})}, & \text{if } m\!\in\!\{D_{k1}\!\cup\!\ldots\!\cup\!D_{kB_k}\}, \\
0, & \text{otherwise},
\end{cases}
\label{eq:power-control-coefficient}
\end{equation}   
where $f(\cdot)$ is a pre-determined function that is calculated locally at AP $m$, and $T_m$ is the set of UEs served by AP$_m$, i.e., given $m$, the set of $k$ for which $m \in \{D_{k1} \cup \ldots \cup D_{kB_k}\}$. $f(\cdot)$ is function of $\mathcal{G}_{mk^\prime}$, a set that comprises long-term statistical knowledge of the channel, such as $\gamma_{mk^\prime}$, or $\beta_{mk^\prime}, k^\prime \in T_m$. 
The denominator in~\eqref{eq:power-control-coefficient} constitutes the normalization term ensuring that~\eqref{eq:power-constraint} is satisfied. 
Importantly, there is no inter/intra-cluster interaction between APs in the selection of the power control coefficients.

The signalling diagram in~\Figref{fig:signaling-diagram} depicts the operation of the proposed framework from the UE $k$ perspective. 
In the figure, we follow the same example as in~\Figref{fig:proposed-system}: UE$_2$, is served by all APs in the cluster $D_1 \cup D_2$, since the user-centric cluster of UE$_2$ involves APs of the cell-centric cluster $D_1$ and $D_2$. 
CPU$_3$ does not participate in serving UE$_2$. 
We recall that the data transmission does not regard only UE $k$. 
A given AP $m$ serve coherently all the UEs belonging to $T_m$, as follows
\begin{equation}
x_m  = \sqrt{\Pd} \sum\nolimits_{k \in T_m} \sqrt{\eta_{mk}} \hat{g}^\ast_{mk} q_k,
\end{equation}
which differs from~\eqref{eq:transmission-model} as not all the UEs are served when a user-centric transmission is designed. 
In the example in~\Figref{fig:signaling-diagram}, it is assumed that the APs selected to form the user-centric cluster are chosen according to the channel quality they offer to UE $k$, e.g., the largest large-scale-fading-based AP selection criterion proposed in~\cite{Ngo2018a}. 

\section{Numerical Results}
Our simulations aim to compare the downlink spectral efficiency (SE) provided by the proposed framework for cell-free massive MIMO with the canonical cell-free massive MIMO, and a conventional multi-cell cooperative MIMO system (e.g., CoMP-JT).
Everything else being equal, the only difference between these three setups consists in the set of APs that serve each single UE.

\subsection{Simulation Scenario}

Instead of implementing a classical wrap-around technique, we use a simple embedding technique to substantially eliminate border effects. We consider square $\mathcal{A}$ of 2.5 km $\times$ 2.5 km. 
In the middle of $\mathcal{A}$, a focus square $\mathcal{B}$ of 1 km $\times$ 1 km is defined. For transmission, all the elements in $\mathcal{A}$ are considered, but only the elements inside of $\mathcal{B}$ are taken into account for the performance evaluation. 
The effect is substantially that elements inside of $\mathcal{B}$ are samples of a ``stationary'' distribution, not affected by edge effects; and that edge effects affect only UEs at the boundary of $\mathcal{A}$. 
We consider 625 single-antenna APs and 125 single-antenna UEs uniformly at random placed in $\mathcal{A}$, such that $M = 100$ APs and $K = 20$ UEs fall into $\mathcal{B}$, and the remaining 525+105 ``dummy'' elements fall into the area between $\mathcal{A}$ and $\mathcal{B}$.

\begin{figure}[!t] \centering
    \subfloat[Example of conventional CoMP-JT.\label{fig:comp}]{
    	\includegraphics[width=.9\columnwidth]{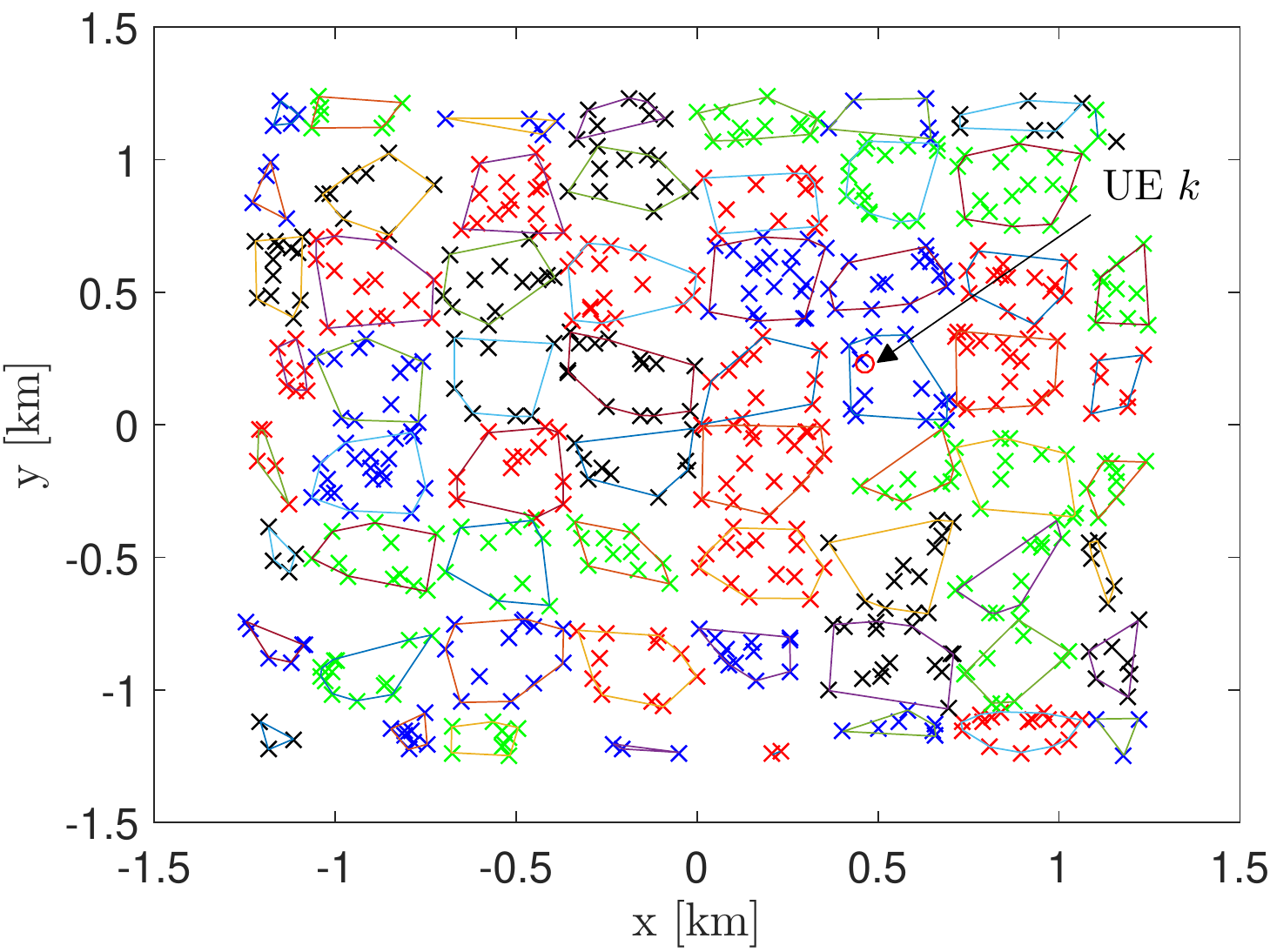}}\hfill
    \subfloat[Example of the proposed cell-free framework with $B_k=2$.\label{fig:proposed}]{
    	\includegraphics[width=.9\columnwidth]{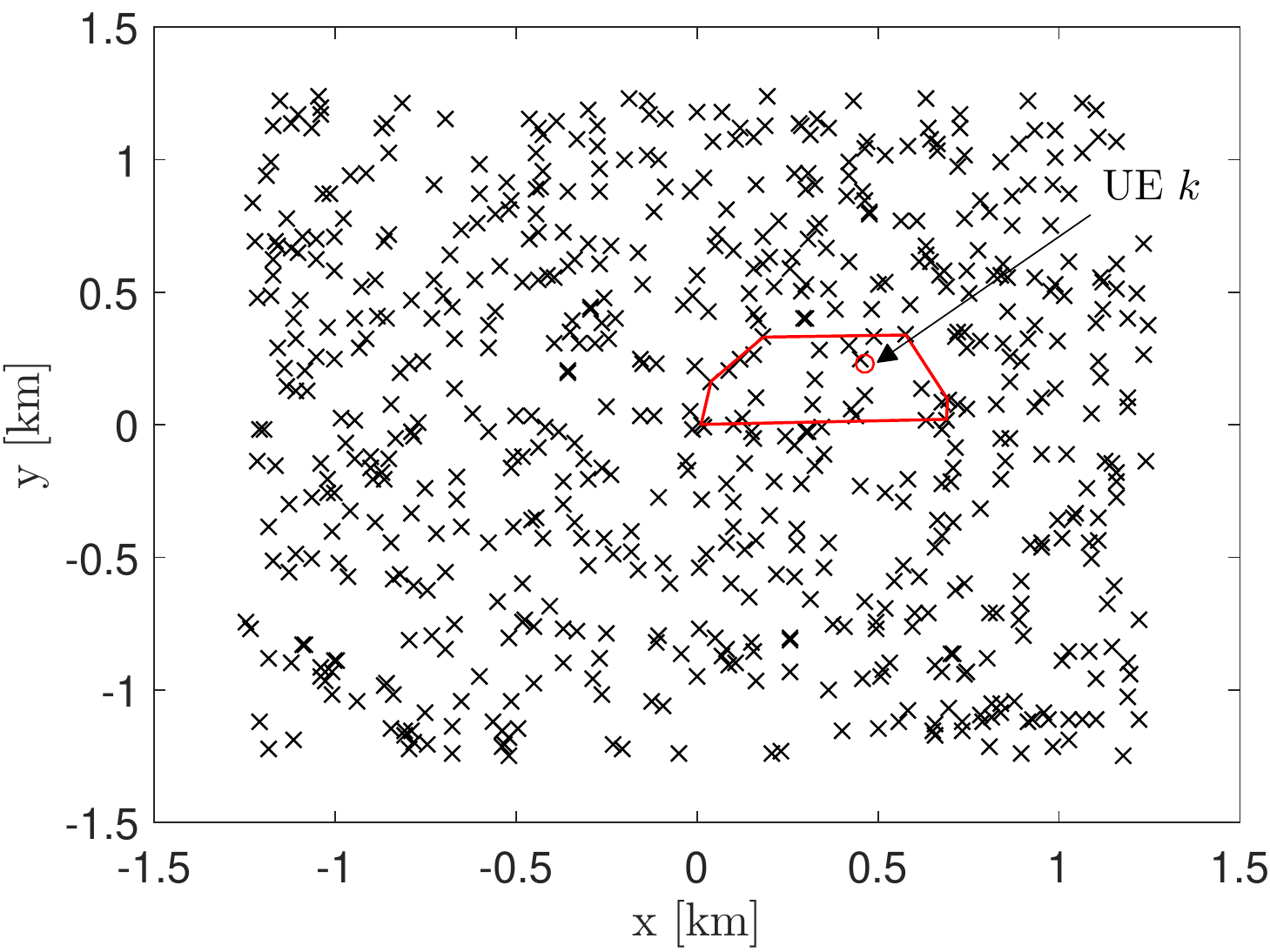}}\hfill
	\caption{Cross markers indicate APs while the circle  marker indicates an arbitrary UE $k$.}
	\label{fig:fig2}  
\end{figure}

Fig.~\subref*{fig:comp} shows the deployment of a CoMP-JT network in the considered area. 
Polygons of different colors contain APs (cross markers) of the same cell-centric cluster, i.e., APs served by the same CPU.
In conventional CoMP-JT, a given UE $k$ (circle marker) is served by all the APs belonging to the cell-centric cluster that guarantees the highest quality service (the cluster containing UE $k$ depicted in Fig.~\subref*{fig:comp}).

The same APs deployment but for the proposed cell-free massive MIMO framework is illustrated in Fig.~\subref*{fig:proposed}. 
In this example, it is assumed that, to serve UE $k$, the APs selected by the user-centric approach belong to two cell-centric clusters ($B_k=2$). Hence, UE $k$ is served by the cell-centric clusters $D_{k1}, D_{k2}$ which in Fig.~\subref*{fig:proposed} have been ``merged'' (they can be identified by visual inspection from Fig.~\subref*{fig:comp}).

Lastly, in canonical cell-free massive MIMO, all the APs in the network serve every single UE (i.e., Fig.~\subref*{fig:comp} with the cluster containing all the APs).

\subsection{Spectral Efficiency Evaluation}

To evaluate the per-user downlink SE, we use the closed-form expression given in~\cite{Ngo2017b}, which assumes single-antenna APs, conjugate beamforming, channel estimation errors and non-orthogonal uplink pilots. 
The channel is modeled as in~\cite{Ngo2017b}, and includes: three-slope path-loss with Hata-COST231 propagation model; uncorrelated shadow fading; and independent Rayleigh fading. The simulation settings also resemble those in~\cite{Ngo2017b}: the maximum radiated power is 200 mW per AP and 100 mW per UE; the carrier frequency is 1.9 GHz, the transmission bandwidth 20 MHz, and the coherence bandwidth 200 kHz; the noise figure is 9 dB; the antenna height of the APs and the UEs are 15 m and 1.65 m, respectively. 

We assume that each AP has a set of 10 orthogonal pilots. The uplink pilots are randomly assigned to the UEs and there is no coordination among the APs in assigning the pilots. 

\begin{figure}[!t]
\centering
\includegraphics[width=\columnwidth]{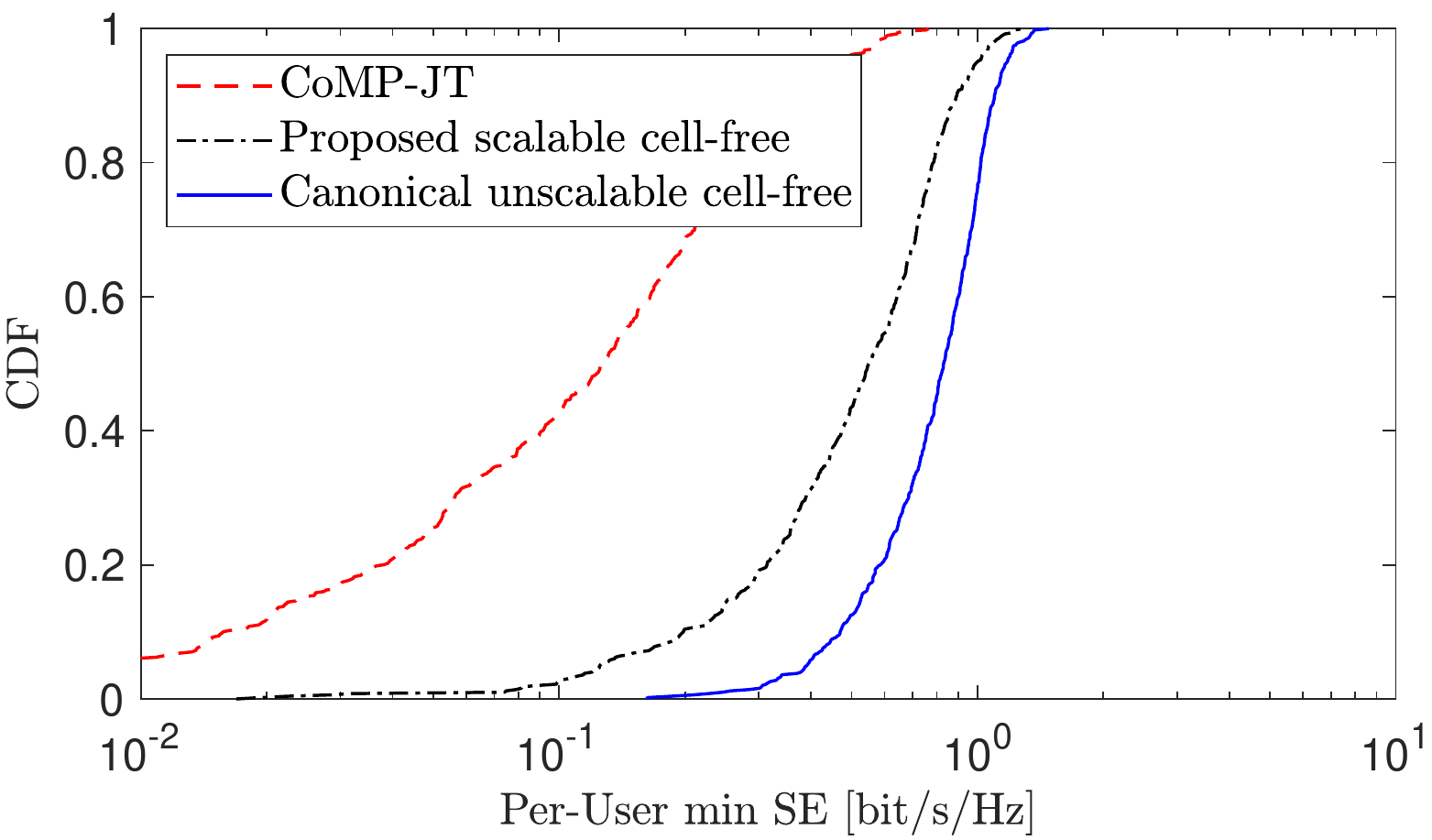}
\caption{The CDF of the downlink per-user minimum spectral efficiency.}
\label{fig:min-se}
\end{figure}

\Figref{fig:min-se} shows the cumulative distribution function (CDF) of the downlink minimum spectral efficiency per UE provided by the proposed scalable cell-free massive MIMO framework, canonical cell-free massive MIMO, and CoMP-JT. In these simulations, we set $\mathcal{G}_{mk} = \gamma_{mk}$, $f(\gamma_{mk}) = 1/\sqrt{\gamma_{mk}}$. 
Hence, the coefficient that adjusts the downlink transmitted power from AP $m$ to UE $k$ is given by 
\begin{equation}
\eta_{mk}\!=\! 
\begin{cases}
\frac{1/{\sqrt{\gamma_{mk}}}}{\sum\limits_{k^\prime \in T_m} \sqrt{\gamma_{mk^\prime}}}, & \text{if } m\!\in\!\{D_{k1}\!\cup\!\ldots\!\cup\!D_{kB_k}\}, \\
0, & \text{otherwise}.
\end{cases}
\label{eq:power-control-coefficient-gamma}
\end{equation}

The AP selection method\footnote{AP selection methods are nothing more than a power control strategy. An AP $m$ is not involved in the transmission to a UE $k$ by setting $\eta_{mk} = 0$.} to form the user-centric clusters is, in this example, based on the AP-to-UE distance: the 5 closest APs are selected and determines the serving cell-centric clusters. 
Results show that the proposed scalable cell-free massive MIMO framework substantially outperforms CoMP-JT. 
This gap derives from the user-centric transmission that suppresses the inter-cell interference. 
Conversely, the performance loss compared to canonical  cell-free massive MIMO is modest. Hence, very few serving cell-centric clusters are sufficient to guarantee good performance.  
Involving more APs in the user-centric cluster by adjusting the AP selection criterion increases the number of serving cell-centric clusters per UE. 
This results in better performance (i.e., the dash-dotted curve approaches the solid one) but also increases the complexity of the system as more APs and CPUs are involved in the coherent transmission and requires coordination.

\subsection{Distributed Power Control Strategies}
The power control strategy chosen in~\eqref{eq:power-control-coefficient-gamma} requires additional motivation.
Firstly, this choice allows to make the power control fully distributed as~\eqref{eq:power-control-coefficient-gamma} involves only local CSI. 
This yields benefits to the scalability of the system as no power control coefficient needs to be exchanged over the front-haul network. 
Secondly, it has low complexity as there is no optimization problem to solve. 
In addition, only long-term channel statistics are used in~\eqref{eq:power-control-coefficient-gamma} thus the updating frequency of the power control coefficients is low (i.e., large-scale fading time scale, which spans over multiple coherence intervals).

\begin{figure}[!t]
\centering
\includegraphics[width=\columnwidth]{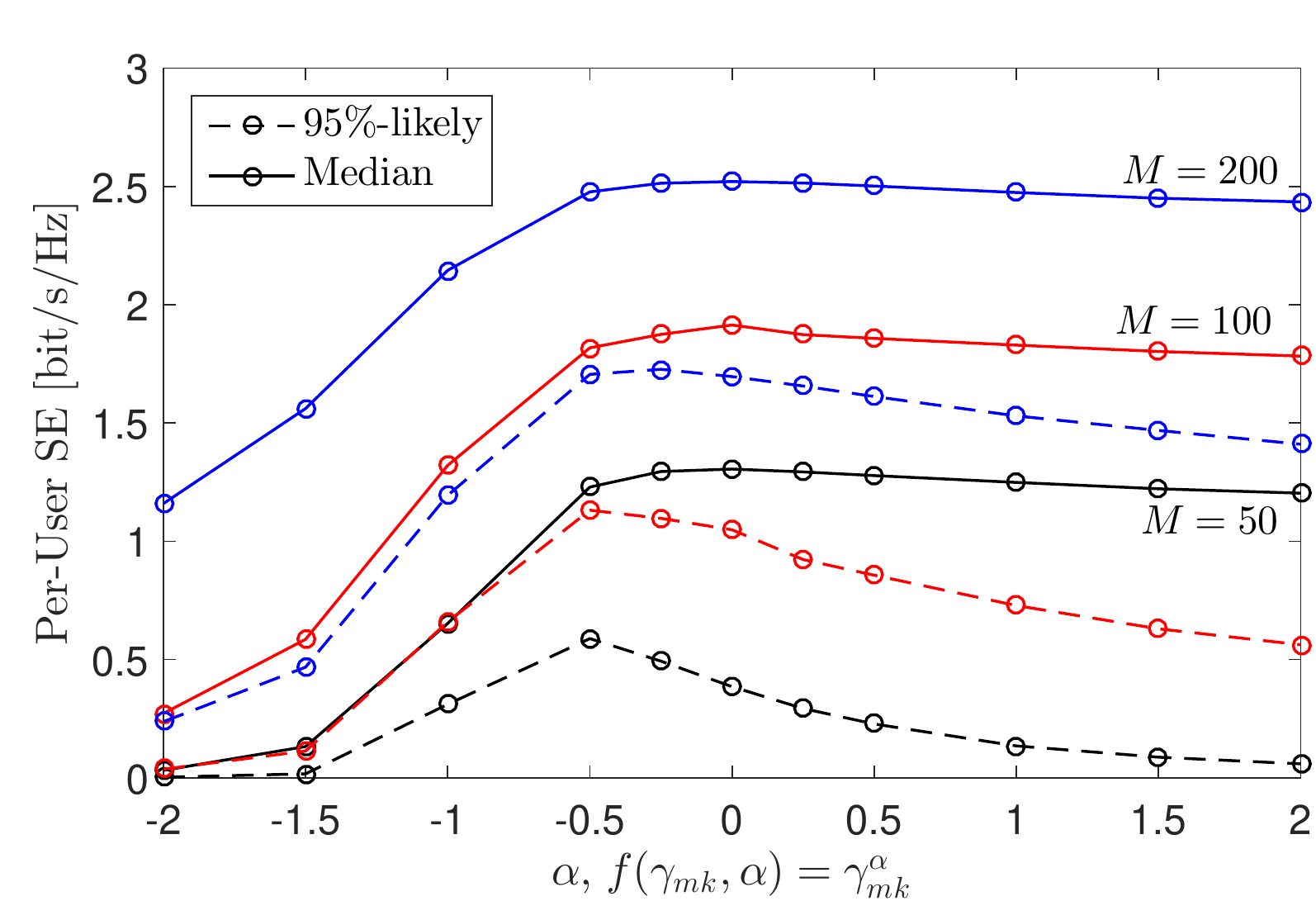}
\caption{Per-user SE versus channel-dependent power control strategy: $\alpha$ adjusts the relationship between $\eta_{mk}$ and $\gamma_{mk}$.}
\label{fig:gamma-plot}
\end{figure}

\Figref{fig:gamma-plot} shows the per-user spectral efficiency achieved by the scalable cell-free massive MIMO framework, varying the exponent $\alpha$ in the function $f(\gamma_{mk},\alpha) = \gamma_{mk}^\alpha$. In these simulations, we use the same settings as in~\Figref{fig:min-se}, and the largest large-scale-fading-based AP selection method, described in~\cite{Ngo2018a}, to form the user-centric clusters.
Results demonstrate that, both in terms of 95\%-likely SE and median SE, very good performance can be achieved by setting $f(\gamma_{mk}) = 1/\sqrt{\gamma_{mk}}$. Increasing the number of APs in the system does not affect the choice of the power control strategy as the user-centric approach selects in any case very few serving cell-centric clusters.
This particular choice of the power control coefficients ensures that the effective power allocated to the service of UE $k$ by AP $m$ is proportional to $\gamma_{mk} \eta_{mk} \propto \sqrt{\gamma_{mk}}$; i.e., the better channel between an AP and a UE, the more power is allocated by the AP to the service of that UE.  This conclusion contrasts with that in~\cite{Nayebi2017a} since the \textit{heuristic uniform power allocation} (HUPA), therein proposed, works well only for large number of serving APs.     

\begin{figure}[!t]
\centering
\includegraphics[width=\columnwidth]{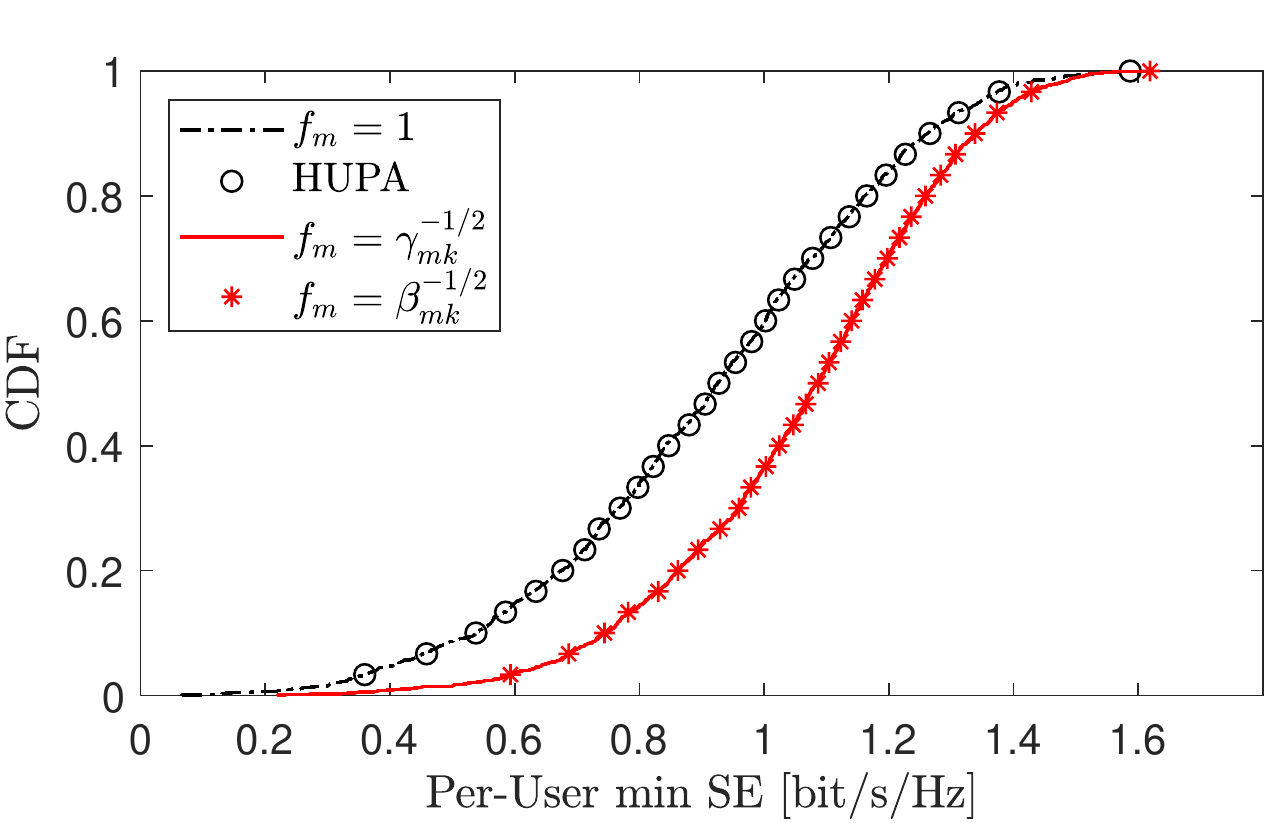}
\caption{CDF of the per-user min SE provided by uniform power control allocation~\cite{Ngo2017b}, HUPA~\cite{Nayebi2017a}, and proposed channel-dependent power allocation with $f(\gamma_{mk}) = 1/\sqrt{\gamma_{mk}}$, and $f(\beta_{mk}) = 1/\sqrt{\beta_{mk}}$.}
\label{fig:power-control}
\end{figure}

The latter is also confirmed by \Figref{fig:power-control} which shows the CDF of the per-user minimum SE achieved by the uniform power allocation in~\cite{Ngo2017b}, the HUPA proposed in~\cite{Nayebi2017a}, and the  proposed channel-dependent power allocation with $f(\gamma_{mk}) = 1/\sqrt{\gamma_{mk}}$.
The proposed power allocation provides substantial SE gain over both the alternative schemes which perform almost identically. This gain derives from the fact that, when user-centric transmission is implemented, very few APs are effectively serving a given UE. 
Under this condition, the exponential behavior of the power scatter plot proposed in~\cite{Nayebi2017a} does not hold. A similar gain is obtained by setting $f(\beta_{mk}) = 1/\sqrt{\beta_{mk}}$, since $\beta_{mk} = \gamma_{mk} + \epsilon_{mk}$~\cite{Ngo2017b}, where $\epsilon_{mk}$ is the variance of the channel estimation error, and $\beta_{mk} = \gamma_{mk}$ if estimation is perfect. 
In general, we can conclude that it is better to ``customize'' the transmitted powers by setting the power control coefficients according to the specific user channel conditions rather than uniforming the power allocation.

\section{Conclusion}
In cell-free massive MIMO, it is  assumed that each UE is being served by  all the APs, which are in turn managed by  a single CPU. 
Such a system is unrealistic and unscalable.

In this study, we proposed a fully distributed and scalable user-centric architecture for cell-free massive MIMO. 
The APs are grouped in cell-centric clusters. Each cluster is managed by a CPU and operates autonomously. 
The presence of multiple CPUs, each one managing a cell-centric cluster, allows for reduced  deployment complexity. 
The UE is then served by all   cell-centric clusters involved in the user-centric cluster. This enables a limited
 distribution of the data payload as the data destined to a given UE is only distributed among the CPUs of the cell-centric clusters selected. 
From the AP perspective,   computational complexity is reduced when it comes to  channel estimation,    calculation of  power control coefficients, and 
precoding/decoding.
From the UE perspective, the   design increases the SE by leveraging the coordination among multiple cell-centric clusters. Very few   clusters need to
cooperate   to achieve performance comparable to the canonical (and unscalable) form of cell-free massive MIMO.

A distributed channel-dependent power control scheme was proposed, where the power control coefficients scale proportionally (at an appropriate rate) with the mean-square of the effective/estimated channel. This policy facilitates fully distributed computation of the power control coefficients.

Future research directions include: derivation of a SE expression that considers the front/back-hauling overhead; performance evaluation under a more realistic channel model that includes channel correlations; and extension to the multi-antenna UE case.

\bibliographystyle{IEEEtran}
\bibliography{IEEEabrv,refs}

\begin{thebibliography}{10}
\providecommand{\url}[1]{#1}
\csname url@samestyle\endcsname
\providecommand{\newblock}{\relax}
\providecommand{\bibinfo}[2]{#2}
\providecommand{\BIBentrySTDinterwordspacing}{\spaceskip=0pt\relax}
\providecommand{\BIBentryALTinterwordstretchfactor}{4}
\providecommand{\BIBentryALTinterwordspacing}{\spaceskip=\fontdimen2\font plus
\BIBentryALTinterwordstretchfactor\fontdimen3\font minus
  \fontdimen4\font\relax}
\providecommand{\BIBforeignlanguage}[2]{{%
\expandafter\ifx\csname l@#1\endcsname\relax
\typeout{** WARNING: IEEEtran.bst: No hyphenation pattern has been}%
\typeout{** loaded for the language `#1'. Using the pattern for}%
\typeout{** the default language instead.}%
\else
\language=\csname l@#1\endcsname
\fi
#2}}
\providecommand{\BIBdecl}{\relax}
\BIBdecl

\bibitem{Zhou2003a}
S.~Zhou, M.~Zhao, X.~Xu, J.~Wang, and Y.~Yao, ``Distributed wireless
  communication system: {A} new architecture for future public wireless
  access,'' \emph{{IEEE} Commun. Mag.}, vol.~41, no.~3, pp. 108--113, Mar.
  2003.

\bibitem{Nayebi2015a}
E.~Nayebi, A.~Ashikhmin, T.~L. Marzetta, and H.~Yang, ``Cell-free {Massive}
  {MIMO} systems,'' in \emph{Proc.~Asilomar Conf. Signals, Syst., Comput.},
  Nov. 2015, pp. 695--699.

\bibitem{Ngo2017b}
H.~Q. Ngo, A.~Ashikhmin, H.~Yang, E.~G. Larsson, and T.~L. Marzetta,
  ``Cell-free {Massive} {MIMO} versus small cells,'' \emph{{IEEE} Trans.
  Wireless Commun.}, vol.~16, no.~3, pp. 1834--1850, Mar. 2017.

\bibitem{Interdonato2018a}
\BIBentryALTinterwordspacing
G.~Interdonato, E.~Bj{\"{o}}rnson, H.~Q. Ngo, P.~Frenger, and E.~G. Larsson,
  ``Ubiquitous cell-free {M}assive {MIMO} communications,'' \emph{CoRR}, vol.
  abs/1804.03421, 2018. [Online]. Available:
  \url{http://arxiv.org/abs/1804.03421}
\BIBentrySTDinterwordspacing

\bibitem{Venkatesan2007a}
S.~Venkatesan, A.~Lozano, and R.~Valenzuela, ``Network {MIMO}: Overcoming
  intercell interference in indoor wireless systems,'' in \emph{Proc.~IEEE
  ACSSC}, 2007, pp. 83--87.

\bibitem{Caire2010b}
G.~Caire, S.~Ramprashad, and H.~Papadopoulos, ``Rethinking network {MIMO}: Cost
  of {CSIT}, performance analysis, and architecture comparisons,'' in
  \emph{Proc.~Information Theory and Applications Workshop (ITA)}, Jan. 2010,
  pp. 1--10.

\bibitem{Feng2013c}
W.~Feng, Y.~Wang, N.~Ge, J.~Lu, and J.~Zhang, ``Virtual {MIMO} in multi-cell
  distributed antenna systems: {C}oordinated transmissions with large-scale
  {CSIT},'' \emph{{IEEE} J. Sel. Areas Commun.}, vol.~31, no.~10, pp.
  2067--2081, Oct. 2013.

\bibitem{Gesbert2010a}
D.~Gesbert, S.~Hanly, H.~Huang, S.~Shamai, O.~Simeone, and W.~Yu, ``Multi-cell
  {MIMO} cooperative networks: A new look at interference,'' \emph{{IEEE} J.
  Sel. Areas Commun.}, vol.~28, no.~9, pp. 1380--1408, Dec. 2010.

\bibitem{Boldi2011a}
M.~Boldi, A.~T{\"o}lli, M.~Olsson, E.~Hardouin, T.~Svensson, F.~Boccardi,
  L.~Thiele, and V.~Jungnickel, ``Coordinated multipoint {(CoMP)} systems,'' in
  \emph{Mobile and Wireless Communications for {IMT}-Advanced and Beyond},
  A.~Osseiran, J.~Monserrat, and W.~Mohr, Eds.\hskip 1em plus 0.5em minus
  0.4em\relax Wiley, 2011, pp. 121--155.

\bibitem{Marsch2011a}
P.~Marsch, S.~Br{\"u}ck, A.~Garavaglia, M.~Schulist, R.~Weber, and A.~Dekorsy,
  ``Clustering,'' in \emph{Coordinated multi-point in mobile communications:
  From theory to practice}, P.~Marsch and G.~Fettweis, Eds.\hskip 1em plus
  0.5em minus 0.4em\relax Cambridge, 2011, ch.~7, pp. 139--159.

\bibitem{irmer2011coordinated}
R.~Irmer, H.~Droste, P.~Marsch, M.~Grieger, G.~Fettweis, S.~Brueck, H.-P.
  Mayer, L.~Thiele, and V.~Jungnickel, ``{Coordinated multipoint: Concepts,
  performance, and field trial results},'' \emph{{IEEE} Commun. Mag.}, vol.~49,
  no.~2, pp. 102--111, 2011.

\bibitem{Lozano2013a}
A.~Lozano, R.~W. Heath, and J.~G. Andrews, ``Fundamental limits of
  cooperation,'' \emph{{IEEE} Trans. Inf. Theory}, vol.~59, no.~9, pp.
  5213--5226, Sep. 2013.

\bibitem{Marzetta2016a}
T.~L. Marzetta, E.~G. Larsson, H.~Yang, and H.~Q. Ngo, \emph{Fundamentals of
  {M}assive {MIMO}}.\hskip 1em plus 0.5em minus 0.4em\relax Cambridge
  University Press, 2016.

\bibitem{Nayebi2017a}
E.~Nayebi, A.~Ashikhmin, T.~L. Marzetta, H.~Yang, and B.~D. Rao, ``Precoding
  and power optimization in cell-free {Massive} {MIMO} systems,'' \emph{{IEEE}
  Trans. Wireless Commun.}, vol.~16, no.~7, pp. 4445--4459, Jul. 2017.

\bibitem{Papadogiannis2008a}
A.~Papadogiannis, D.~Gesbert, and E.~Hardouin, ``A dynamic clustering approach
  in wireless networks with multi-cell cooperative processing,'' in
  \emph{Proc.~IEEE International Conference on Communications (ICC)}, May 2008,
  pp. 4033--4037.

\bibitem{Baracca2012b}
P.~Baracca, F.~Boccardi, and V.~Braun, ``A dynamic joint clustering scheduling
  algorithm for downlink {CoMP} systems with limited {CSI},'' in
  \emph{Proc.~IEEE International Symposium on Wireless Commun. Systems
  (ISWCS)}, Aug. 2012, pp. 830--834.

\bibitem{Jungnickel2014b}
V.~Jungnickel, K.~Manolakis, W.~Zirwas, B.~Panzner, V.~Braun, M.~Lossow,
  M.~Sternad, R.~Apelfrojd, and T.~Svensson, ``The role of small cells,
  coordinated multipoint, and massive {MIMO} in {5G},'' \emph{{IEEE} Commun.
  Mag.}, vol.~52, no.~5, pp. 44--51, May 2014.

\bibitem{Pan2018b}
C.~Pan, M.~Elkashlan, J.~Wang, J.~Yuan, and L.~Hanzo, ``User-centric {C-RAN}
  architecture for ultra-dense {5G} networks: {C}hallenges and methodologies,''
  \emph{{IEEE} Commun. Mag.}, vol.~56, no.~6, pp. 14--20, Jun. 2018.

\bibitem{Yuan2017c}
J.~Yuan, S.~Jin, W.~Xu, W.~Tan, M.~Matthaiou, and K.~Wong, ``User-centric
  networking for dense {C-RANs}: {H}igh-{SNR} capacity analysis and antenna
  selection,'' \emph{{IEEE} Trans. Commun.}, vol.~65, no.~11, pp. 5067--5080,
  Nov. 2017.

\bibitem{Bjornson2013d}
E.~Bj{\"{o}}rnson and E.~Jorswieck, ``Optimal resource allocation in
  coordinated multi-cell systems,'' \emph{Foundations and Trends in
  Communications and Information Theory}, vol.~9, no. 2-3, pp. 113--381, 2013.

\bibitem{Ngo2018a}
H.~Q. Ngo, L.~N. Tran, T.~Q. Duong, M.~Matthaiou, and E.~G. Larsson, ``On the
  total energy efficiency of cell-free {M}assive {MIMO},'' \emph{{IEEE} Trans.
  Green Commun. Netw.,}, vol.~2, no.~1, pp. 25--39, Mar. 2018.

\end{thebibliography}
\end{document}